\documentclass[a4paper,12pt,reqno]{amsart}
\usepackage{amsfonts,amssymb,amsmath}
\newtheorem{theorem}{Theorem} 
 
\newtheorem{lemma}[theorem]{Lemma} 
\newtheorem{proposition}[theorem]{Proposition} 
\newtheorem{corollary}[theorem]{Corollary} 
\newtheorem{remark}[theorem]{Remark}

\DeclareMathOperator{\expect}{{\mathbb E}}

\newcommand{\eps}{\varepsilon}

\newcommand{\R}{\mathbb{R}}
\newcommand{\Z}{\mathbb{Z}}
\newcommand{\I}{\mathcal{I}}
\newcommand{\A}{\mathcal{A}}

\newcommand{\X}{\mathcal{X}}
\newcommand{\W}{\mathcal{W}}
\newcommand{\F}{\mathcal{F}}

\begin{document}
\title[Gibbs measures on Wiener paths]
{Gibbs measures for \\self-interacting Wiener paths}
\author{Massimiliano Gubinelli} 
\address{Dip. di Matematical Applicata ``U.Dini'' -- Universit\`a di
Pisa\\ via Bonanno 25b\\
56125 Pisa - Italia.}
\date{March 2006, rev.~1}
%\maketitle

\begin{abstract}
In this note we study a class of specifications over $d$-dimensional
Wiener measure which are invariant under uniform translation of
the paths. This degeneracy is removed by restricting the measure to the
$\sigma$-algebra generated by the increments of the coordinate process.
We address the problem of existence and uniqueness of Gibbs measures
 and prove a  central limit theorem for the rescaled increments. 
These results apply to the study of the ground state
 of the Nelson model of a quantum particle
interacting with a scalar boson field.\\[0.3cm]
\end{abstract}

\maketitle
\section{Introduction}
The theory of Gibbs measures for lattice spin models or continuous
point processes is by now a well established subject of probability
theory. References on the subject are the book of Georgii~\cite{Georgii}, the early
monograph by Preston~\cite{Preston}, the readable and concise introduction of
F\"ollmer~\cite{Follmer} and the pedagogical review by van~Enter et~al.~\cite{jsp}. 

Gibbs measures on  path spaces
are a more unexplored domain and only recently a series of works
started a systematic study of a class of Gibbs measures on  paths 
motivated by applications in Quantum
Mechanics~\cite{BLS,BHLMS02,BetzLor03,Betz03}. However, part of these results
are obtained through functional analytic approaches which does not help
to fully understand the probabilistic structure of these models.

An interesting class of Gibbs measures is obtained by
perturbing the Wiener measure $\W$ on $C(\R,\R^d)$ by the exponential
of a (finite-volume)  energy of the form
$$
H_T(x) = \int_{-T}^T V(x_t) dt + \int_{-T}^T dt \int_{-T}^T ds W(x_t,x_s,t-s)
$$
where $x$ is the path and the functions $V$ and $W$ are 
interpreted as interactions potentials. In this way we obtain
finite-volume measures $\mu_T$ given by
\begin{equation}
%  \label{eq:3}
 \mu_{\lambda,T}(dx) = \frac{e^{-\lambda H_T(x)}}{Z_T} \W(dx) 
\end{equation}
The study of these measures in the limit $T \to \infty$ has been
addressed in the works cited above and a series of conditions on $V$
and $W$ have been found which are sufficient for the existence of the
limit (in the topology of local weak convergence) and for its
uniqueness.  

Here we are interested in a class of models whose energy function enjoy
an invariance under shift of the paths $x \to x + c$ where $c$ is a
fixed vector in $\R^d$. In particular we take $V = 0$ and let
$W(\xi,\xi',t) = W(\xi-\xi',t)$ so that our finite-volume energy reads
$$
H_T(x) =  \int_{-T}^T dt \int_{-T}^T ds\, W(x_t-x_s,t-s).
$$

A relevant model for which we will prove existence of a unique Gibbs measure for small coupling is the ($d=3$, UV regularized) Nelson model
corresponding to the interaction energy given by the function
\begin{equation}
  \label{eq:nelson-3d}
W(\xi,t) = -\frac{1}{1+|\xi|^2+|t|^2}.  
\end{equation}

Nelson model is a member of a wide class of potentials justified by applications to Quantum Mechanics  which are in the form
\begin{equation}
  \label{eq:spectral}
W(\xi,t)  = \int_{\R^d} \frac{dk}{2\omega(k)} |\rho(k)|^2 e^{- i k\cdot \xi - \omega(k) t}
\end{equation}
where $\omega(k) : \R^d \to \R_+$ is the dispersion law of a scalar boson field. Nelson model corresponds to the case $\omega(k) = |k|$ and $\rho(k)$ with a fast decay at infinity (ultraviolet  cutoff) and $|\rho(0)| > 0$ (no infrared cutoff).

In~\cite{BS03} Betz and Spohn prove that if $\rho$ satisfy the following integrability conditions:
\begin{equation}
  \label{eq:bs-cond}
\int_{\R^d} dk |\rho(k)|^2 
 \left( \omega(k)^{-1} + \omega(k)^{-2} + \omega(k)^{-3}\right) < \infty  
\end{equation}
then
 the rescaled coordinate process $X^{(\eps)}_t =
\eps^{-1/2} X_{t/\eps}$ weakly converge as $\eps\to 0$ to a
$d$-dimensional Brownian motion. Note that this result does not cover the $d=3$ Nelson model since eq.~(\ref{eq:bs-cond}) is not satisfied. Moreover they proved that if
$$
\int_{\R^d} dk |\rho(k)|^2 |k|^2
 \left(  \omega(k)^{-2} + \omega(k)^{-4}\right) < \infty,
$$
the limiting Brownian motion has a non-trivial diffusion constant (i.e. different from zero). Their approach is based on the use of
an auxiliary Gaussian field which allow to ``linearize'' the
interaction and to see the process $X$ as the projection of a Markov
process on a larger state space. The functional central limit theorem
 then follows by using a technique due to  Kipnis and Varadhan.

In~\cite{Spohn87} it is pointed out that this class of
models can be naturally recast as models of spins on $\Z$ with single
spin space $C([0,1],\R^d)$. This is obtained by cutting the path into
pieces and considering each piece as a spin. In this
representation the interaction becomes multi-body and under natural
conditions on the time decay of $W$, this multi-body interaction is
long range with a power decay. The main problem with this approach is that the Wiener measure does not factorizes appropriately after such decomposition.
Moreover the shift invariance of the Hamiltonian prevents to have tightness of
the family of measures $\{\mu_{\lambda,T}\}_{T > 0}$ (e.g. when pinned at $\{-T,T\}$). 

The natural solution to both problems
 is 
to consider \emph{increments} of the Wiener path as the basic
variables. No relevant information is lost on the (local
or global) behavior  of the path.  Since increments over disjoint intervals are independent, the reference measure now  factorizes over the product of countably many independent degrees of freedom. Techniques form the theory of one
dimensional spin systems can be successfully applied. This approach  will give also estimates
on the decay of correlation and on the strong mixing coefficients of the Gibbs measures. 
Then the central limit theorem (CLT)
can be proved using standard results on strongly mixing processes.
However we should point out that our conditions for the CLT does \emph{not} cover the 3d Nelson model.

\subsection*{
Plan of the paper} In Sec.~\ref{sect:increments} we introduce the
space of increments and the reference Gaussian measure induced by the
Wiener measure on the paths. In Sec.~\ref{sec:existence} we prove
existence of the Gibbs measures under very mild conditions and in
Sec.~\ref{sec:uniqueness} we give some more restrictive conditions
under which uniqueness can be proved. Sec.~\ref{sec:lower} is devoted to the proof of uniform lower bounds for the diffusion constant under the Gibbs measures.
Finally in Sec.~\ref{sec:diffusive} we prove the CLT for a suitable
class of interactions.

\section{Brownian increments}
\label{sect:increments}
Fixed a finite union of intervals $I \subseteq \R$ consider the linear subspace $\X_I \subset C(I^2,\R^d)$ such that $x \in \X_I$
iff $x_{tt} = 0$ for any $t \in I$ and satisfy the cocycle condition
\begin{equation}
  \label{eq:cocycle}
 x_{su} + x_{ut} = x_{st}, \qquad s,t,u \in I 
\end{equation}
Let $\X := \X_\R$ and $\X_T := \X_{[-T,T]}$ for each $T > 0$.

%  define the norm
% $$
% \|x\|_{\gamma,T} := \sup_{s,t \in [-T,T]} \frac{|X_{st}|}{|s-t|^\gamma}
% $$
% and note that if $\|x\|_\gamma < \infty$ this implies that both
% mappings $t \mapsto x_{st}$ and  $s \mapsto x_{st}$ are H\"older
% continuous of exponent $\gamma > 0$. 
% The couple $(\X_T,\|\cdot\|_{\gamma,T})$ is a Banach space. 

On $\X$ consider the Gaussian measure $\mu$ such that, if $\{X_{st}\}_{s,t \in\R}$ is
the coordinate process:
$$
\expect^\mu [X_{st }]= 0, \qquad \expect^\mu [X_{st}^2] = |s-t|
$$
and $X_{st},X_{uv}$ are independent iff $(s,t)$ and $(u,v)$ are
disjoint intervals.
This process can be easily understood as deriving from a $d$-dimensional Brownian
motion $B$ by setting $X_{st} = B_t - B_s$. 
Take $a \le b$ and let $\mathcal{F}_{[a,b]} = \sigma(X_{st} : a \le s \le t \le b)$ the
$\sigma$-field generated by the increments in the interval $[a,b]$ and
let
 $\mathcal{F}^{T,c}_{[a,b]} = \sigma(X_{st} : [s,t] \subset [-T,T] \backslash [a,b])$. In general, for a finite union of intervals $I \subset \R$ we let $\F_I = \sigma(X_{ts}: [t,s] \subseteq I)$.
Given an interval $I$ and two paths $x,y \in \X$ we let 
 $z = (x \otimes_I y)$ the
 unique path in $\X$ such that $z|_{I\times
  I} = x |_{I\times  I}$ and $z|_{I^c\times
  I^c} = y |_{I^c\times  I^c}$. Translations $\{\tau_a : a\in\R\}$ acts on $\X$ in the canonical way: $(\tau_a x)_{st} = x_{s+a,t+a}$.

In the following $C$ will stay for any positive constant, not necessarily the same from line to line and not depending on anything else unless otherwise stated.

\section{Gibbs measures}
\label{sec:existence}
The specification $\Pi$ is the family of proper probability kernels
$\pi_I(\cdot|\cdot)$ with $I$ running on the set $\I$ of all intervals of $\R$ such
that
\begin{equation}
  \begin{split}
\pi_I(dx|y) = Z_{I}^{-1}(y) e^{-\lambda U_{I}(x) -\lambda V_{I}(x,y)} \mu_I\times \delta_{I^c,y}(dx)   
  \end{split}
\end{equation}
for any $I \in \I$,
where $\delta_{I^c,y}$ is the Dirac measure on $\X_{I^c}$
concentrated on the path $y|_{I^c\times I^c}$, $U_I$ is an $\F_I$
measurable function and $V_I$ is $\F_I \times \F_{I^c}$ measurable
given by
\begin{equation}
\label{eq:def-U}
U_I(x) = \int_{I\times I} dt  ds \, W(x_{st},t-s)  
\end{equation}
\begin{equation}
\label{eq:def-V}
V_I(x,y) = \int_{J(I)} dt ds \, [W((x \otimes_I y)_{st},t-s) -W((0 \otimes_I y)_{st},t-s) ] 
\end{equation}
where $W \in C(\R^d\times \R ; \R)$ and
\begin{equation}
  \label{eq:cone}
J(I) = (I^+ \times I^-) \cup (I^- \times I^+) \cup (I \times I^c) \cup (I^c \times I)
\end{equation}
with $I^+$ and $I^-$ the positive and negative half lines cut off by
$I$ ($I^c = I^- \cup I^+$). Note that $J(I)$ is the ``cone'' of
influence of the increments in the interval $I$. That is, increments
outside $J(I)$ are independent from increments inside $I$ and any
increment inside $J(I)$ can be written as a sum of an increment in $I$
and increments outside $J(I)$.

\begin{remark}
  The specification $\Pi$ is translation invariant
  (cfr. Preston\cite{Preston}, page 48), i.e. for any $a\in\R$ and any
  $I \in \mathcal{I}$ we have $\pi_I(\tau_a f| \tau_a y) =
  \pi_{\tau_a(I)}(f|y)$ where $\tau_a$ acts on functions as $\tau_a f
  (x) = f(\tau_a(x))$ and on intervals as $\tau_a([b,c]) = [b+a,c+a]$.
\end{remark}

\begin{remark}
The definition~(\ref{eq:def-V}) of the interaction between the increments in $I$ and outside $I$ is justified by the fact that only relative energies matter in the specification $\Pi$. As we will shortly see, we can allow for models where the irrelevant constant we subtracted can be infinite.  
\end{remark}

For $x \in \X$, $\xi \in \R^d$, $a \ge 0$ define
$$
Q(x,\xi,a) := \int_{-\infty}^0 dt \int_{0}^\infty ds \, |W(\xi+x_{st},a+s-t)-W(x_{st},a+s-t)|.
$$

On $W$ we will make the following assumptions:
\begin{itemize}
\item[\textbf{(H1)}] $\sup_{x \in \X} |U_I(x)| \le C |I|^2$ and
$
\sup_{x \in \X}\int_{-\infty}^\infty dt \, |W(x_{0t},|t|)| < \infty
$;
\item[\textbf{(H2)}]
For $\xi\in \R^d$,
$
\sup_{a \ge 0} \sup_{x \in \X} Q_T(x,\xi,a) \le C  (1+|\xi|).
$
\end{itemize}    

\begin{lemma}
\label{lemma:good-pi}
Assume $(H1)$ and $(H2)$, then the specification $\Pi$ is well defined and
\begin{equation}
  \label{eq:bound-V}
  |V_{[a,b]}(x,y)| \le C (1+|b-a|+|x_{ab}|) 
\end{equation}
for any $a<b$ and all $x,y \in \X$.
\end{lemma}
\begin{proof}
Let $I = [a,b]$ for some $a<b$.
By hypothesis $(H1)$ $U_I(x)$ is well under control so we have only to care about $V_I$. According to eq.~(\ref{eq:cone}) the cone $J(I)$ can be split in two regions $R_1 = I^+ \times I^- \cup I^- \times I^+$ and $R_2 = I \times I^c \cup I^c \times I$. The second requirement of hypothesis $(H1)$ is enough to show that
\begin{equation*}
  \int_{R_2} dt ds \, [W((x \otimes_I y)_{st},t-s) -W((0 \otimes_I y)_{st},t-s) ] \le C |I|.
\end{equation*}
For the region $R_1$ we proceed as follows. Let $[a,b]=I$ and consider the integral
\begin{equation*}
  \begin{split}
\mathcal{J} & = \int_{I^+\times I^-} dt ds\, [W((x \otimes_I y)_{st},t-s) -W((0 \otimes_I y)_{st},t-s) ]  
\\ & = \int^{+\infty}_b dt \int_{-\infty}^a ds \, [W((x \otimes_I y)_{st},t-s) -W((0 \otimes_I y)_{st},t-s) ]  
\\ & = \int^{+\infty}_b dt \int_{-\infty}^a ds \, [W(x_{ab}+y_{tb}+y_{sa},t-s) -W(y_{tb}+y_{sa},t-s) ]  
  \end{split}
\end{equation*}
where we used the cocycle property to write $(x \otimes_I y)$ as a sum of increments.
By a change of variables we obtain
\begin{equation*}
  \begin{split}
\mathcal{J} = &\int^{+\infty}_0 dt \int_{-\infty}^0 ds \, [W(x_{ab}+y_{b+t,b}+y_{a+s,a},t-s+b-a)
\\ & \qquad \qquad  -W(y_{b+t,b}+y_{a+s,a},t-s+b-a) ]  
  \end{split}
\end{equation*}
and now it is not difficult to check that $\mathcal{J} = Q(\tilde y, x_{ab},b-a)$ for some $\tilde y \in \X$ so the integral $\mathcal{J}$ is well defined under hypothesis $(H2)$. Applying the same argument for the set $I^-\times I^+$ and summarizing what we have found we end up with the bound~(\ref{eq:bound-V}).
\end{proof}

\begin{theorem}
\label{th:existence}
Assume $(H1)$ and $(H2)$, then there exists at least one translation invariant Gibbs measure
for the specification $\Pi$.  
\end{theorem}
\begin{proof}
Existence will follow from Theorem~3.1 of Preston~\cite{Preston} once we have shown that the specification $\Pi$ satisfy the following two conditions (referred respectively as $(3.11)$ and $(3.8)$ in the given reference):
\begin{itemize}
\item[a)] \textit{(uniform control)} for every interval $I$ there exists a finite measure $\omega_I$  on $\F_I$ and another interval $K$ such that the kernels $\{\pi_K(\cdot|y)|_{\F_I} : y \in \X\}$ are uniformly absolutely continuous w.r.t $\omega_I$;
\item[b)] \textit{(quasilocality)} given some local function $f$ (i.e. $f \in \F_I$ for some interval $I$), some interval $K$ and $\eps > 0$ then there exists an interval $K'$ and a $\F_{K'}$ measurable function $g_{K'}$ such that $|\pi_K(f|y)-g_{K'}(y)| < \eps$ for any $y \in \X$.
\end{itemize}

Let us check $(a)$.
Fix $y \in \X$ and take an interval $K$ such that $I \subset K$. For $f \in \mathcal{F}_I$, $f \ge 0$ we have
\begin{equation}
  \begin{split}
\pi_K(f|y) & = \frac{\int_{\X_K} f(x ) e^{-\lambda U_K(x)
    -\lambda V_K(x,y)} \mu_K(dx)}{\int_{\X_K}  e^{-\lambda U_K(x)
    -\lambda V_K(x,y)} \mu_K(dx)}
\\  & = \frac{\int_{\X_H} \mu_H(dz) \int_{\X_I} \mu_I(dx) f(x) e^{-\lambda U_K(x \otimes_I z)
    -\lambda V_K(x \otimes_I z,  y)}}{\int_{\X_H} \mu_H(dz) \int_{\X_I} \mu_I(dx) e^{-\lambda U_K(x \otimes_I z)
    -\lambda V_K(x \otimes_I z,   y)}}   
  \end{split}
\end{equation}
where $H = K \backslash I$.
From Lemma~\ref{lemma:good-pi} and setting $K = [a,b]$ we have the following uniform bound
\begin{equation}
|V_K(x \otimes_I z,y)| \le C  (1+|K|+|(x \otimes_I z)_{ab}|).
\end{equation}
The expression on the r.h.s. can be bounded using the inequality
$$
|(x \otimes_I z)_{ab}| = |z_{ac}+x_{cd}+z_{db}| \le |z_{ac}| + |z_{db}| + |x_{cd}|
$$ 
where we set $I = [c,d]$ so that $a < c < d < b$ and used the cocycle property.
Moreover $|U_K(x)| \le C |K|^2$
so, for some constant $M_K$, we obtain
\begin{equation*}
  \begin{split}
\pi_K(f|y) & \le M_K \frac{\int_{\X_H} \mu_H(dz) \int_{\X_I} \mu_I(dx) f(x) e^{\lambda C|x_{cd}| + \lambda C (|z_{ac}|+|z_{db}|)}}{\int_{\X_H} \mu_H(dz) \int_{\X_I} \mu_I(dx) e^{-\lambda C|x_{cd}| - \lambda C (|z_{ac}|+|z_{db}|)}} 
\\ & \le M_K \frac{\int_{\X_H} \mu_H(dz) e^{\lambda C (|z_{ac}|+|z_{db}|)} \int_{\X_I} \mu_I(dx) f(x) e^{\lambda C|x_{cd}|}}{\int_{\X_H} \mu_H(dz) e^{- \lambda C (|z_{ac}|+|z_{db}|)} \int_{\X_I} \mu_I(dx) e^{-\lambda C|x_{cd}|}}     
  \end{split}
\end{equation*}
Under $\mu_K(dz)$ the r.v. $z_{ac}$ and $z_{db}$ have Gaussian distribution so their modulus is exponentially integrable. The denominator is strictly greater than zero by Jensen's inequality and the integrability of the modulus, so setting
\begin{equation}
  \label{eq:omega}
\omega_I(dx) := \frac{e^{\lambda C|x_{cd}|} \mu_I(dx)}{\int_{\X_I} e^{\lambda C|x_{cd}|} \mu_I(dx)} \end{equation}
we have that there exists a constant $M'_K$ such that $\pi_K(f|y) \le M'_K \omega_I(f)$ for any $f \in \F_I$ and any $y \in \X$, proving property $(a)$. 

Let us now turn to quasilocality.
Given $I,K,f$ as in $(b)$, for any interval $K'$ such that $K \subset K'$ define
$$
g_{K'}(y) := \frac{\int_{X_K} \mu_K(dx) f(x \otimes_K y) e^{-\lambda U_K(x) -\lambda V_{K}^{K'}(x,y)} }{
\int_{X_K} \mu_K(dx) f(x \otimes_K y) e^{-\lambda U_K(x) -\lambda V_{K}^{K'}(x,y)}}
$$
where
\begin{equation}
\label{eq:V-approx}
V_K^{K'}(x,y) := \int_{J(K)\cap K'} dt ds \, [W((x \otimes_I y)_{st},t-s)-W((0 \otimes_I y)_{st},t-s)].  
\end{equation}
Observe that, by definition, $g_{K'}$ is a $\F_{K'}$ measurable function and by $(H2)$ and arguments like those used in lemma~\ref{lemma:good-pi} we see that $V_{K}^{K'}(x,y)$ converges uniformly to $V_K(x,y)$ as $K' \uparrow \R$. Then $g_{K'}(y)$ is $\F_{K' \cup I}$ measurable and $g_{K'}(y) \to \pi_K(f|y)$ uniformly as $K' \uparrow \R$ (cfr. Preston~\cite{Preston}, Prop.~5.3). This proves $(b)$. 
So we can apply Thm 3.7 of Preston and conclude that the set of Gibbs measures for the specification $\Pi$ is non-empty.

Moreover conditions $(a)$ and $(b)$ and the translation invariance of the specification $\Pi$ are also enough to apply Thm. 4.3 of Preston which ensures that there exists at least one Gibbs measure which is invariant under the countable abelian group of rational translations $\{\tau_a : a \in \mathbb{Q}\}$. This is enough to conclude.
\end{proof}

\begin{remark}
The hypothesis $(H2)$ above is more general than to require the potential to
be \emph{absolutely summable} (see~\cite{Georgii})
, which would be equivalent to have
$$
\int_{-\infty}^{0}ds  \int_{0}^{\infty} dt\, |W(x_{st},t-s)| < \infty
$$  
uniformly for $x \in \X$. The Nelson model given by eq.~(\ref{eq:nelson-3d}) is an example of non-absolutely summable potential for which is necessary to consider relative energies.
\end{remark}

\begin{remark}
Nelson model
 satisfies assumptions $(H1)$ and $(H2)$. In particular,
for $(H2)$  we have
\begin{equation}
\nabla_\xi Q(x,\xi,a) =  2 \int_{-\infty}^0 dt \int_{0}^\infty ds \,
 \frac{1}{(1+|\xi+x_{st}|^2+|a+s-t|^2)^2} (\xi+x_{st})
\end{equation}
so that
\begin{equation}
  \begin{split}
|\nabla_\xi Q(x,\xi,a)| & \le  2 \int_{-\infty}^0 dt \int_{0}^\infty ds \,
 \frac{1}{(1+|\xi+x_{st}|^2+|a+s-t|^2)^{3/2}} 
\\ & \le  2 \int_{-\infty}^0 dt \int_{0}^\infty ds \,
 \frac{1}{(1+|a+s-t|^2)^{3/2}}     
\\ & \le  C (1+|a|)^{-1}
  \end{split}
\end{equation}

\end{remark}

%\subsection{Some variations}

Define the measure
\begin{equation}
\mu_{\lambda,T}(dx) := Z_{\lambda,T}^{-1} e^{-\lambda H_T(x)} \mu(dx)   
\end{equation}
on $\X$
with
$$
H_T(x) := \int_{-T}^T dt \int_{-T}^T ds \, W(x_{st},s-t)
$$

% Let  $\tilde \mu_{\lambda,T}$ is the unique measure on $\X$ which coincide with $\mu_{\lambda,T}$ on $\F_{[-T,T]}$ and that is $\delta_{0}$ on $\F_{[-T,T]^c}$.

\begin{corollary}
\label{prop:tight}
Assume $(H1)$ and $(H2)$. Then
the family of measures $\{\mu_{\lambda,T} \}_{T \ge 1}$ is tight (for
the topology of local convergence). Any cluster point is a Gibbs
measure for the specification $\Pi$.
\end{corollary}
\begin{proof}
Consider the approximate kernels $\pi^T_I$ given by
\begin{equation}
  \begin{split}
\pi^T_I(dx|y) = Z_{I,T}^{-1}(y) e^{-\lambda U_{I}(x) -\lambda V^T_{I}(x,y)} \mu_I\times \delta_{[-T,T]\backslash I,y}(dx)   
  \end{split}
\end{equation}
for any interval $I \subseteq[-T,T]$, where $V^T_I = V_I^{[-T,T]}$ (recall eq.~(\ref{eq:V-approx})). Then for $f \in \F_I$  with $I \subseteq [-T,T]$ we have
\begin{equation}
\label{eq:comp}
  \begin{split}
\mu_{\lambda,T}(f) = \int_{\X}  \pi^T_K(f|y)  \mu_{\lambda,T}(dy)    
  \end{split}
\end{equation}
for any interval $K \subseteq [-T,T]$.
The kernels $\pi^T_K$ can be controlled like the kernels $\pi_K$  with the measure $\omega_I$ defined in eq.~(\ref{eq:omega}), i.e. $\pi^T_K(f|y) \le M_K' \omega_I(f)$ for some constant $M'_K$ depending on $K$. Then same is  true for the measures $\mu_{\lambda,T}$  showing the tightness. The continuity of the function $W$ and the absolute convergence of the integrals defining the potentials $U$ and $V$ ensures that the functions $y \mapsto \pi_I^T(f|y)$ and $y \mapsto \pi_I(f|y)$ are continuous for any continuous bounded local $f$, moreover $\pi_I^T(f|y) \to \pi_I(f|y)$ as $T \to \infty$ uniformly in $y$. Then passing to the limit in eq.~(\ref{eq:comp})  we get that any accumulation point $\mu_\lambda$ of the family $\{\mu_{\lambda,T}\}_T$ satisfy the equation $\mu_\lambda(f) = \mu_\lambda (\pi_{K}(f|\cdot))$ for any interval $K$ and any continuous bounded local $f$. This implies that $\mu_\lambda$ is a Gibbs measure for the specification $\Pi$.
\end{proof}

\section{Uniqueness}
\label{sec:uniqueness}
In the rest of this paper we will assume always that conditions $(H1)$ and $(H2)$ holds. 
 A straightforward condition for uniqueness of the Gibbs measure is given by
\begin{itemize}
\item[\textbf{(H3)}]
$$
\int_0^\infty dt \int^0_{-\infty} ds \,|W(x_{st},t-s)-W(0,t-s)| \le C
$$
uniformly in $x \in \X$.
\end{itemize}  
Note that $(H3)$ implies $(H2)$.
\begin{proposition}
\label{prop:unique1}
Assuming $(H1)$ and $(H3)$ there exists a unique Gibbs measure for the
specification $\Pi$.  
\end{proposition}
\begin{proof}
Observe that under $(H3)$ and for any interval $I$ we have a uniform bound $|V_I(x,y)| \le C_V$ independent of $I$. Indeed the cone $J(I)$ can be written as the (non-disjoint) union of the sets $(I^-)^c \times I^-$, $I^- \times (I^-)^c$, $I^+ \times (I^+)^c$, $(I^+)^c \times I^+$ and for each of these sets w the double integral in the definition of $V_I$ is bounded uniformly thanks to $(H3)$.

Then for any bounded $f \in \F_I$ with $f \ge 0$
\begin{equation}
  \begin{split}
\pi_I(f|y) & = \frac{\int_{\X_I} f(x ) e^{-\lambda U_I(x)
    -\lambda V_I(x,y)} \mu_I(dx)}{\int_{\X_I}  e^{-\lambda U_I(x)
    -\lambda V_I(x,y)} \mu_I(dx)}
%\\  &
 \le e^{2\lambda C_V} \frac{\int_{\X_I} f(x ) e^{-\lambda U_I(x)
    } \mu_I(dx)}{\int_{\X_I}  e^{-\lambda U_I(x)
  } \mu_I(dx)}
  \end{split}
\end{equation}
so $\pi_I(f|y) \le e^{2\lambda C_V} \widetilde\omega_I(f)$ where
$$
\widetilde\omega_I(dx) :=  \frac{   e^{-\lambda U_I(x)}  \mu_I(dx)}{ \int_{\X_I}  e^{-\lambda U_I(x) }\mu_I(dx)}
$$
Moreover we can also estimate $\pi_I(f|y)$ from below as $\pi_I(f|y) \ge e^{-2\lambda C_V} \widetilde\omega_I(f)$.
Together these bounds  imply:
\begin{equation}
%  \label{eq:1}
  \pi_I(f|y)  \ge e^{-4 \lambda C_V}
\pi_I(f|z)
\end{equation}
uniformly for any couple $z,y \in \X$ and any interval $I$. This, according to a theorem of Georgii~\cite{Georgii}
(Prop. 8.38) implies uniqueness of the Gibbs measure.
\end{proof}

\begin{remark}
Condition $(H3)$ is satisfied whenever
$$
\sup_{\xi} |W(\xi,t)| \le C (1+|t|)^{-(2+\delta)}
$$
for some $\delta > 0$ or when
$$
W(\xi,t) = C (1+|\xi|^2+|t|^2)^{-(3/2+\delta)}
$$  
always for some $\delta > 0$.
\end{remark}

%\medskip

\subsection{Mapping to a discrete model}
For any size $L > 0$ the space $\X$ splits into a product space of
countably many copies of $\X_L = C([0,L]^2,\R^d)$ as follows.
Introduce the sequence on intervals $\tau_i = [iL,(i+1)L]$, $i \in
\mathbb{Z}$ and for $x \in \X$ let $x_i = x|_{\tau_i \times \tau_i}$
considered as an element of $\X_L$. Let $F : \X \to \X_L^{\Z}$ be the
map $(F x)(i) = x_i$ for any $i \in \Z$ and $x \in \X$. Note that it
is well defined the inverse map $F^{-1} : \X_L^{\mathbb{Z}} \to \X$
and that through $F$ we can identify $\X \simeq \X_L^{\Z}$. Moreover
using $F^{-1}$ the measure $\mu$ factorizes accordingly : $\mu =
F^{-1}_* (\otimes_{n \in \Z} \mu_L^{(n)})$ where each $\mu^{(n)}_L$ is
the Wiener measure over increments in the interval $\tau_n$. Using
this mapping the path measure can be seen as a one-dimensional
(unbounded) lattice spin system with single spin space $\X_L$. 

In particular we can see the total energy $W_I(x|y) = U_I(x) + V_I(x,y)$ in a bounded interval
as originating from potentials $U_{ij}$ such that
$$
U_{ij}(x) = \int_{iL}^{(i+1)L} dt \int_{jL}^{(j+1)L} ds\, W(x_{ts},t-s)
$$
in the following way:
$
W_I(x|y) =\sum_{i,j : I \cap I_{ij} \neq \emptyset} U_{ij}(x \otimes_I y)
$
where $I_{ij}$ is the smallest closed interval containing the set
$\{iL,(i+1)L,jL,(j+1)L\}$. 

Introduce condition $(H3b)$:
\begin{itemize}
\item[\textbf{(H3b)}] There exists  constants $C<\infty$ and $\gamma > 2$ such that
$$
\sup_{\xi\in \R^d}|W(\xi,t)| \le C (1+|t|)^{-\gamma}.
$$
\end{itemize}

Let $\pi_{I,[-n,n]}(\cdot|\cdot)$ be the following probability kernels
$
\pi_{I,[-n,n]}(\cdot|y) = \pi_{[-n,n]}(\cdot|y)|_{\F_I}
%\int_{\X_{[-n,n]}} \pi_I(f| z \otimes_{[-n,n]} y) \mu_{[-n,n]}(dz)
$.
Note that the map $y \mapsto \pi_{I,[-n,n]}(\cdot|y)$ is
$\F_{[-n,n]^c}$ measurable.

\begin{theorem}
\label{th:dobr2}
Under condition $(H3b)$ there exist a unique Gibbs measure $\mu_\lambda$ for the
specification $\Pi$. Moreover for any interval $I=[a,b]$, the marginal distribution
$\mu_{\lambda,I}$ on $\F_I$, satisfy
\begin{equation}
  \label{eq:corr-bounds-xx}
\expect^\mu \left|\frac{d\mu_{\lambda,I}}{d\mu_I} - \frac{d\pi_{I,[-n,n]}(\cdot|y)}{d\mu_I}
\right| \le  \chi(n-\max(|a|,|b|))
\end{equation}
uniformly in $y \in \X$
where $\chi(n)$ is a decreasing function which can be chosen to be
$\chi(n) = A' |\log n|^{3-\gamma} |n|^{2-\gamma}$ for some constant $A'>0$.
\end{theorem}
\begin{proof}
The theorem follows directly from Thm.~1 in~\cite{dobr-uni} (whose proof is contained in~\cite{dobr3}). Indeed
it is easy to check that under condition $(H3b)$  
 the discretized model (e.g. with $L=1$) fulfills all the
required hypotheses.
In particular, setting $M_U(k) = \sup_{|i-j| \ge k} \sup_x
|U_{ij}(x)|$,  it holds that
$
\sum_{k \ge 0} k M_U(k) < \infty  
$
and  there exists a non-increasing function $\psi(n) = A
|n|^{2-\gamma}$ such that
$
\sum_{k \ge n} k M_U(k)\le \psi(n) 
$.  So, according to this result we can choose the function $\chi(n)$
to be $\chi(n) = A' |\log n|^{3-\gamma} |n|^{2-\gamma}$.
\end{proof}

\subsection{The contraction technique}
In a more general setup than conditions $(H3)$ or $(H3b)$ we are able to prove
 the uniqueness of the Gibbs measure in the small coupling regime
(i.e. when $\lambda$ is small) using the contraction technique
introduced by Dobrushin~\cite{Dobr1,Dobr2}.   

Let $\|\cdot\|$ denote the sup norm on $\X_L$ and $\|\cdot\|_1$ the
Lipschitz semi-norm on $C(\X_L,\R)$: 
\begin{equation}
 \|f\|_1 := \sup_{x\neq y \in \X_L} \frac{|f(x)-f(y)|}{\|x-y\|} 
\end{equation}

Given two probability
measures $\mu$, $\nu$ on $\X_L$ define the
Kantorovich-Rubinstein-Vasherstein (KRV) distance
$$
d(\mu,\nu) := \sup_{f} \frac{|\mu(f)-\nu(f)|}{\|f\|_1}
$$
where the sup is taken for $f \in C(\X_L,\R)$.

Let $C(\X)$ the space of continuous functions on $\X$ which are
uniform limit of bounded local functions (i.e. depending only on
finitely many $\X_L$ factors).
For $f \in C(\X)$ let
\begin{equation}
  \label{eq:lip1}
\|f\|_{1,i} :=  \sup \left\{ \frac{|f(x)-f(y)|}{\|x_i-y_i\|} :
 x,y \in\X  \quad \text{$ x_j=y_j$ for $j \neq i$} \right\}  
\end{equation}
and let $\text{Lip}(\X)$ the class of functions $f \in C(\X)$ which satisfy
$$
|f(x)-f(y)| \le \sum_i \|x_i-y_i\| \|f\|_{1,i}, \qquad \sum_i
\|f\|_{1,i} < \infty 
$$

Let $\pi_i := \pi_{\tau_i}$. Define the Dobrushin interaction matrix $C$ as
\begin{equation}
C_{ik} :=
\sup\left\{\left. \frac{d(\pi_{i}(\cdot|y),\pi_{i}(\cdot|z))}{\|y_k-z_k\|}
\right | \; y,z \in\X :  \text{$ y_j=z_j$ for $j \neq k$} \right\}  
\end{equation}

A \emph{tempered measure} $\nu$ is a measure on $\X$ for which  there
exists $z \in \X$ such that
$$
\sup_i \int_{\X} \|x_i-z_i\| \nu(dx) < \infty
$$

Note that, as a by product of the proof of Thm.~\ref{th:existence} we
have that, under conditions $(H1)$ and $(H2)$ there exists tempered  Gibbs
measures for the specification $\Pi$.

The following result is originally due to Dobrushin~\cite{Dobr1,
Dobr2}. The formulation in terms of the KRV metric is taken from
F\"ollmer~\cite{Follmer} (see also~\cite{kuensch}) where interested readers can find the relative proofs.

\begin{theorem}[Dobrushin's uniqueness theorem]
Whenever
\begin{equation}
  \label{eq:uniq-cond}
\sup_k \lim_{n \to \infty} \sum_{i} (C^n)_{ik} = 0  
\end{equation} 
there exists a
unique tempered Gibbs measure $\mu_\lambda$ for the specification $\Pi$.   
\end{theorem}

\begin{remark}
Note that a sufficient condition for eq.~(\ref{eq:uniq-cond}) is given
by $\sup_k \sum_{i} C_{ik} < 1$.   
\end{remark}

Next we will apply these general results to our model.
Let us introduce another class of potentials which is more general
than $(H3)$:
\begin{itemize}
\item[\textbf{(H4)}] For some $K(W) <\infty$ and some $\alpha > 3$ we have
$$
\sup_{\xi \in \R^d} |\nabla_\xi \nabla_\xi W(\xi,t)  |   \le \frac{K(W)}{(1+|t|)^{\alpha}}.
$$
\end{itemize}

\begin{theorem}
\label{th:dobr}
Under hypothesis $(H4)$   we have
\begin{equation}
\label{eq:boundCthm}
C_{ij} \le C  \lambda K(W)  \sigma^2 
   (1+|i-j|)^{2-\alpha}.  
\end{equation}
for some constant $C<\infty$ and some $\sigma^2 < \infty$.
Therefore, for small $\lambda$ there
exists a unique tempered Gibbs measure $\mu_\lambda$ for the
specification $\Pi$.
\end{theorem}
\begin{proof}
We want to check that the specification $\Pi$ satisfy the requirements
  for the application of Dobrushin's uniqueness criterion~(\ref{eq:uniq-cond}).

  Take $y,z \in \X$ and let
$y^r = y + r (z-y)$ for $r\in [0,1]$. Then use the fundamental theorem of calculus to write
\begin{equation*}
  \begin{split}
 \pi_i(f|z)-\pi_i(f|y) & = 
\int_0^1 dr \partial_r \left[Z_i(y^r)^{-1} \int_{\X_L}
  f(x \otimes_i y) e^{-\lambda U_i(x) -\lambda V_i(x,y^r)} \mu_i(dx)
  \right]
\\ & + \left[Z_i(z)^{-1} \int_{\X_L}
  [f(x \otimes_i z)-f(x \otimes_i y)] e^{-\lambda U_i(x) -\lambda
    V_i(x,z)} \mu_i(dx) \right]
\\ & = \int_0^1 dr A(r) + B
   \end{split}
\end{equation*}
where we let
$$
Z_i(y) := \int_{\X_L} e^{-\lambda U_i(x) -\lambda V_i(x,y)} \mu_i(dx).
$$

The $B$ term is estimated by 
$
 |B| \le  \sum_{k\neq i} \|f\|_{1,k} \|z_k-y_k\| 
$
which is finite if $f \in \text{Lip}(\X)$. 

Let $J_i = J(\tau_i)$ and $J_{ij} = J_i
\cap J_j$. For simplicity take first $y,z \in \X$ such that $y=z$ outside $\tau_j$ and compute
$$
|\partial_r (V_i(x,y^r)-V_i(w,y^r))| \le \int_{J_{ij}} dt ds \sup_\xi
|\nabla_\xi \nabla_\xi W(\xi,t-s)|\, \|x_i-w_i\| \|y_j-z_j\|
$$ 
If $i \neq j$ this integral contains only contributions with time span
greater than $L|i-j-1|$ so if we assume that
$$
\sup_\xi |\nabla_\xi\nabla_\xi  W(\xi,t)| \le K(W) (1+|t|)^{-\alpha}
$$
we get
$$
|\partial_r (V_i(x,y^r)-V_i(w,y^r))| \le C K(W)  (1+|i-j|)^{2-\alpha}
\|x_i-w_i\| \|y_j-z_j\|. 
$$ 

For general $y,z \in \X$ we have
\begin{equation}
  \label{eq:estimate-V}
  |\partial_r (V_i(x,y^r)-V_i(w,y^r))| \le C K(W) \sum_{j \neq i} (1+|i-j|)^{2-\alpha} \|x_i-w_i\| \|y_j-z_j\|. 
\end{equation}

And with $\pi_i(\cdot|y^r) = \nu^r$
\begin{equation}
  \begin{split}
A(r) & = \partial_r \int f(\cdot \otimes_i y) d\nu^r 
%\\ &
= \partial_r \left[Z_i(y^r)^{-1} \int_{\X_L}
  f(x \otimes_i y) e^{-\lambda U_i(x) -\lambda V_i(x,y^r)} \mu_i(dx)
  \right]
\\ &=  \int_{\X} (f(x\otimes_i y)-f(w \otimes_i y)) \left[ \lambda \partial_r
  V_i(x\otimes_i y^r) - \partial_r \log Z_i(y^r) \right] \nu^r(dx) 
  \end{split}
\end{equation}
for any $w \in \X_L$.

This implies
\begin{equation}
\label{eq:estimate-A}
  \begin{split}
|A(r)|^2 & \le  \int_{\X} (f(x\otimes_i y)-f(w \otimes_i y))^2  \nu^r(dx)
 \text{Var}_{\nu^r}\left[  \lambda \partial_r
  V_i(\cdot \otimes_i y^r )\right]
\\ & \le   \|f\|_{1,i}^2  \int_{\X} \|x_i-w\|^2  \nu^r(dx)
 \text{Var}_{\nu^r}\left[  \lambda \partial_r
  V_i(\cdot \otimes_i y^r )\right]   
\\ & \le   \|f\|_{1,i}^2  \sigma_i^2 
 \text{Var}_{\nu^r}\left[  \lambda \partial_r
  V_i(\cdot \otimes_i y^r )\right]   
  \end{split}
\end{equation}
where
$
\sigma_i^2 := \sup_{y\in\X} \inf_{w \in \X_L} \int_{\X} \|x_i-w\|^2  
\pi_i(dx|y).
$
Eq.~(\ref{eq:estimate-V}) implies
$$
\left\{\text{Var}_{\nu^r}\left[  \lambda \partial_r
  V_i(\cdot \otimes_i y^r )\right]\right\}^{1/2}   \le 
C K(W)  \sigma_i \sum_j
  \lambda (1+|i-j|)^{2-\alpha}  \|y_j-z_j\|
$$
and using this estimate into eq.~(\ref{eq:estimate-A}) we get
\begin{equation*}
  \left|A(r) \right| \le C K(W) \|f\|_{1,i}  \sigma_i^2 \sum_j
  \lambda (1+|i-j|)^{2-\alpha}  \|y_j-z_j\|.
\end{equation*}
Then
\begin{equation}
\label{eq:lip-bound}
  \begin{split}
 \left|\pi_i(f|y)-\pi_i(f|z)\right| & \le 
 C \lambda K(W) \|f\|_{1,i}  \sigma_i^2
  \sum_{j \neq i} (1+|i-j|)^{2-\alpha}  \|y_j-z_j\| 
\\ & \qquad + 
 \sum_{j\neq i} \|f\|_{1,j} \|z_j-y_j\| 
  \end{split}
\end{equation}
According to this bound, if $f \in \text{Lip}(\X)$ we have also $\pi_i(f|\cdot) \in
\text{Lip}(\X)$ provided $\alpha > 3$.

For $f(x) = f(x_i)$ the bound~(\ref{eq:lip-bound}) reads
\begin{equation*}
  \begin{split}
 \left|\pi_i(f|y)-\pi_i(f|z)\right| & \le 
 C \lambda K(W)  \|f\|_{1,i}  \sigma_i^2 \sum_{j \neq i} 
   (1+|i-j|)^{2-\alpha}  \|y_j-z_j\| 
  \end{split}
\end{equation*}
which means that
\begin{equation}
d(\pi_i(\cdot|y),\pi_i(\cdot|z)) \le C\lambda K(W)   \sigma_i^2   \sum_{j \neq i} 
   (1+|i-j|)^{2-\alpha}  \|y_j-z_j\|. 
\end{equation}
Moreover if $y=z$ outside $\tau_j$ (i.e. if $y|_{\tau_k} = z|_{\tau_k}$ for any $k \neq j$) then 
\begin{equation}
d(\pi_i(\cdot|y),\pi_i(\cdot|z)) \le C \lambda K(W) \sigma_i^2    
   (1+|i-j|)^{2-\alpha}  \|y_j-z_j\|  
\end{equation}
which in turn implies
$
%\label{eq:boundC}
C_{ij} \le C  \lambda K(W)  \sigma_i^2 
   (1+|i-j|)^{2-\alpha}.  
$
So provided 
$
%  \label{eq:2}
\sigma = \sup_i \sigma_i < \infty 
$
and $\alpha > 3$
we have that $\sum_i C_{ij} \le C' \sigma^2 \lambda K(W) L^{-1}$ and for $\lambda$ sufficiently small we obtain $\sup_j \sum_{i} C_{ij} < 1$ and by applying Dobrushin criterion we can conclude that
 there exists a unique
tempered Gibbs measure $\mu_\lambda$ associated to the specification $\Pi$.
The condition $\sigma^2 < \infty$ follows easily from $(H1)$ and $(H2)$ and the computations of Thm.~\ref{th:existence}.
\end{proof}

\begin{remark}
In the case of the Nelson model we have
$$
|\nabla_\xi \nabla_\xi W(\xi,t)| \le C {(1+|t|^2)^{-2}}
$$  
uniformly in $\xi\in\R^d$, so $\alpha = 4$ and we can apply the above result.
\end{remark}

\section{Lower bound on the diffusion constant}
\label{sec:lower}

Here we provide a lower bound for the second moment of the increments under the Gibbs measure. This will be useful below in the proof of the CLT.

\begin{theorem}
\label{th:lower}
Under condition $(H4)$
 there exists a positive constant $\sigma^2_->0$ such that
$$
\expect^{\mu_{\lambda,T}} [X_{ab}^2 ] \ge \sigma^2_- |a-b|
$$
uniformly in $T,a,b$.
\end{theorem}
\begin{proof}
Integration by parts on $\mu$ is given by the formula
\begin{equation*}
 \expect^{\mu} [X_{ab} F] = \int_{a}^b dt \, \expect^{\mu}[ D_t F ] 
\end{equation*}
where $D_t$ is the Malliavin derivative (see e.g.~\cite{nualart}). This, of course, when both sides make sense.
By integration by parts  we have for
disjoint intervals $[a,b]$, $[c,d]$ with $b < c$:
\begin{equation}
  \begin{split}
\expect^{\mu_{\lambda,T}} [X_{ab} X_{cd}] & =  -
 \expect^{\mu_{\lambda,T}} [A_{ab} X_{cd}]      
%\\ & 
=
\expect^{\mu_{\lambda,T}} [- B_{ab,cd} + A_{ab} A_{cd}]      
  \end{split}
\end{equation}
with
$$
A_{ab} = \int_{a}^{b}ds\, \lambda D_{s}(H_T(X))
%$$
\quad\text{and}\quad
%$$
B_{ab,cd} = \int_{a}^{b}ds \int_{c}^{d}dt\, \lambda D_t D_{s}(H_T(X)).
$$
While
\begin{equation}
\label{eq:ibp-1}
  \begin{split}
\expect^{\mu_{\lambda,T}} [X_{ab}^2 ] & = |a-b|  -
 \expect^{\mu_{\lambda,T}} [A_{ab} X_{ab}]      
%= |a-b| -  \expect^{\mu_{\lambda,T}} [ B_{ab,ab}] + \expect^{\mu_{\lambda,T}} [ A_{ab}^2]      
  \end{split}
\end{equation}
Apply Cauchy-Schwartz inequality to $\expect^{\mu_{\lambda,T}} [A_{ab} X_{ab}]  $ to get the lower bound
\begin{equation}
\label{eq:ineq-1}
  \expect^{\mu_{\lambda,T}} [X_{ab}^2 ] \ge 
|a-b|  -
 \left(\expect^{\mu_{\lambda,T}} [A_{ab}^2] \right)^{1/2}
\left(\expect^{\mu_{\lambda,T}}[ X_{ab}^2]\right)^{1/2}.
\end{equation}
One more integration by parts on the r.h.s of eq.~(\ref{eq:ibp-1}) gives
\begin{equation}
\label{eq:ibp-2}
  \begin{split}
\expect^{\mu_{\lambda,T}} [X_{ab}^2 ] &       
= |a-b| -  \expect^{\mu_{\lambda,T}} [ B_{ab,ab}] + \expect^{\mu_{\lambda,T}} [ A_{ab}^2]      
  \end{split}
\end{equation}
which gives another  inequality
\begin{equation*}
\expect^{\mu_{\lambda,T}} [ A_{ab}^2] \le \expect^{\mu_{\lambda,T}} [X_{ab}^2 ] +  \expect^{\mu_{\lambda,T}} [ |B_{ab,ab}|]. 
\end{equation*}
Use the fact that $|x+y|^{1/2} \le |x|^{1/2} + |y|^{1/2}$ to get
\begin{equation}
\label{eq:ineq-2}
\left(\expect^{\mu_{\lambda,T}} [ A_{ab}^2]\right)^{1/2} \le \left(\expect^{\mu_{\lambda,T}} [X_{ab}^2 ]\right)^{1/2} + \left( \expect^{\mu_{\lambda,T}} [ |B_{ab,ab}|]\right)^{1/2}. 
\end{equation}
Combining together eq.~(\ref{eq:ineq-1}) and eq.~(\ref{eq:ineq-2}) we get the lower bound
\begin{equation}
\label{eq:bound-fin-1}
 2 \expect^{\mu_{\lambda,T}} [X_{ab}^2 ] \ge 
|a-b| - \left( \expect^{\mu_{\lambda,T}} [ |B_{ab,ab}|]\right)^{1/2} 
\left(\expect^{\mu_{\lambda,T}}[ X_{ab}^2]\right)^{1/2}. 
\end{equation}

If we compute the Malliavin derivatives, we get:
\begin{equation}
  \begin{split}
  D_{t} (H_T(X)) & = 2 \int_{-T}^T du \int_{u}^T dv \, D_t 
  W(X_{uv},u-v)    
\\ & = 2 \int_{-T}^{t} du \int_{t}^{T} dv \,
  W_{x}(X_{uv},u-v)    
  \end{split}
\end{equation}
and
\begin{equation}
  \begin{split}
  D_{t} D_{s}(H_T(X)) & = 2 \int_{-T}^T du \int_{u}^T dv \, D_t D_s
  W(X_{uv},u-v)    
\\ & = 2 \int_{-T}^{s} du \int_{t}^{T} dv \,
  W_{xx}(X_{uv},u-v)    
  \end{split}
\end{equation}
where $W_{x}(\xi,t) = \nabla_\xi W(\xi,t)$ and $W_{xx}(\xi,t) = \nabla_\xi \nabla_\xi W(\xi,t)$.
%It is straightforward to see that these equations pass to the limit
%when $T \to \infty$.

Using condition $(H4)$ we get
\begin{equation}
  \begin{split}
|  D_{t} D_{s}(H_T(X))| & \le
 2 \int_{-T}^{s} du \int_{t}^{T} dv \, \frac{C}{(1+|u-v|)^{\alpha}}
 \le
 C (1+|s-t|)^{2-\alpha}.
   \end{split}
\end{equation}
And since $\alpha > 3$ we obtain the bound
$
 \expect^{\mu_{\lambda,T}} [ |B_{ab,ab}|] \le C\lambda  |a-b|
$
uniformly in $T$.

Using this last estimate in the inequality~(\ref{eq:bound-fin-1}) above we get
\begin{equation*}
 2 \expect^{\mu_{\lambda,T}} [X_{ab}^2 ] \ge 
|a-b| - (C\lambda)^{1/2}|a-b|^{1/2} 
\left(\expect^{\mu_{\lambda,T}}[ X_{ab}^2]\right)^{1/2} 
\end{equation*}
Calling $y = \left(\expect^{\mu_{\lambda,T}}[ X_{ab}^2]\right)^{1/2} |a-b|^{-1/2}$ we have
$
 2 y^2 -1 + (C\lambda)^{1/2}y \ge 0
$
with $y \ge 0$. This implies that $y \ge \sigma_-$ for some $\sigma_-> 0$ and we have obtained the lower bound
$
\expect^{\mu_{\lambda,T}} [X_{ab}^2 ] \ge \sigma^2_- |a-b|
$.
\end{proof}

\begin{remark}
Using
$
2|ab| \le |a|^2+|b|^2
$
we have
\begin{equation*}
  \begin{split}
\expect^{\mu_{\lambda,T}} [X_{ab}^2] & \le  |a-b| + \frac{1}{2}
 \expect^{\mu_{\lambda,T}} [X_{ab}^2]      
 + \frac{1}{2} \expect^{\mu_{\lambda,T}} \left[A_{ab}^2 \right]      
  \end{split}
\end{equation*}
and then
$\expect^{\mu_{\lambda,T}} [X_{ab}^2]  \le  2 |a-b| + \expect^{\mu_{\lambda,T}} \left[A_{ab}^2\right]      
$.
This equation and the decay of correlations can provide upper bounds for the diffusion constant.  
\end{remark}

\section{Diffusive behavior}
\label{sec:diffusive}
According to Thms.~\ref{th:dobr2} and~\ref{th:dobr} under condition $(H3b)$ or $(H4)$
(the latter in the small coupling regime)  we have a unique 
Gibbs measure  which we denote $\mu_\lambda$ as above and for which we
have polynomial decay of correlations.  
Then we address the problem of  the long-time behavior of the increment
process under diffusive rescaling. 
Thm.~\ref{th:lower} rules out the possibility of sub-diffusive behavior of the paths
whenever the interaction decays fast enough. This holds irrespective of the magnitude of the coupling constant $\lambda$ and of the character of the potential (attractive or repulsive).
The next theorem establishes that (under some more restrictive
assumption than those needed to obtain uniqueness of the Gibbs measure) we
actually have diffusive behavior of the increment process.

\begin{theorem} Assume $(H4)$ holds. Then in either of the following two situations:
  \begin{itemize}
\item[(a)] condition $(H3b)$ with $\gamma > 3$, or
  \item[(b)]  small $\lambda$ and $\alpha >
  4$;
  \end{itemize}
 the r.v. $X^{\eps}_{t,s} = \eps^{1/2} X_{\eps^{-1}t,\eps^{-1}s}$ weakly
converges to an isotropic
Gaussian vector as $\eps \to 0$.   
\end{theorem}

\begin{proof}
Consider the random variables $Y_i = \langle v, X_{Li,L(i+1)}\rangle
\in \F_{\tau_i}$ for some fixed vector $v \in \R^d$
and for $\Lambda \subset \Z$ let $\A_\Lambda = \sigma(X_i, i \in
\Lambda)$. 
If $\Lambda_1, \Lambda_2 \subset \Z$ let
$d(\Lambda_1,\Lambda_2) = \inf( |n-k|, n \in \Lambda_1, k \in
\Lambda_2)$. Define the following mixing coefficients for the measure
$\mu_\lambda$:
\begin{equation*}
\alpha_{l,k}(n) = \sup\{|\mu_\lambda(A_1 \cap A_2) -\mu_\lambda(A_1)
\mu_\lambda(A_2)| : A_i \in \A_{\Lambda_{i}}, |\Lambda_1| \le k,
|\Lambda_2| \le l, d(\Lambda_1,\Lambda_2) \ge n\}
\end{equation*}
and
\begin{equation*}
\rho(n) = \sup\{|\mathrm{Cov}_{\mu_\lambda}(Z_1,Z_2)| : Z_i \in
L^2(\mu_\lambda,\A_{\{k_i\}}), \|Y_i\|_2 \le 1, |k_1-k_2| \ge n\}  
\end{equation*}
where $\|\cdot\|_p$ denote the $L^p$ norm with respect to the measure
$\mu_\lambda$.

Then, according to a theorem of Bolthausen~\cite{Bolth}, if the
following two conditions holds
\begin{equation}
  \label{eq:cond}
\|Y_i\|_{2+\delta} < \infty, \qquad
\sum_{m=1}^\infty \alpha_{2,\infty}^{\delta/(2+\delta)}(m) < \infty  
\end{equation}
for some $\delta > 0$ and if
\begin{equation}
  \label{eq:cov-b}
\sigma^2 = \sum_{n} \mathrm{Cov}_{\mu_\lambda}(Y_0, Y_n) > 0 
\end{equation}
then the r.v. $S_{n} = (\sigma^2 n)^{-1/2}\sum_{0 \le k
\le n} Y_k$ converges to a standard Gaussian r.v. 

So it will be enough to check the two conditions~(\ref{eq:cond})
and~(\ref{eq:cov-b}). 
For eq.~(\ref{eq:cov-b}) note that
$
\sigma^2  = \lim_{n \to \infty} n^{-1}\expect_{\mu_\lambda} (X_{Ln,0})^2 
$
and by the bound proved in Thm.~\ref{th:lower} we have that this
quantity is bounded below by $\sigma^2_- Ln$ with $\sigma^2_- > 0$, so that we can conclude
$\sigma^2 > 0$.

As for condition~(\ref{eq:cond}) note that each $Y_i$ has moments of
any order due to the fact that the expectation on the measure
$\mu_\lambda$ can be bounded above by expectation on the measure
$\omega_{[0,L]}$ defined in the proof of Thm.~\ref{th:existence} which
can the be directly estimated and shown to be finite. 

Under assumption $(a)$ we are in the conditions to apply Thm.~\ref{th:dobr2} and the inequality~(\ref{eq:corr-bounds-xx}) is enough to
prove that $\alpha_{2,\infty}(n) \le C \chi(n) = C' |n|^{2-\gamma}$. Then for $\gamma > 3$ condition~(\ref{eq:cond}) can be satisfied by choosing $\delta$ sufficiently large.

It remains to check that the mixing coefficient $\alpha(n)$ is
sufficiently summable in case $(b)$. A technical difficulty is that the contraction coefficients for the KRV metric are not suitable to estimate the strong mixing coefficients $\alpha_{l,k}(m)$. This
because we cannot uniformly approximate the indicator functions (needed to estimate probabilities) with
functions in $\mathrm{Lip}(\X)$. However this technical difficulty can
be easily overcome by using a slightly different norm in the Dobrushin
contraction technique.
Let $\|f\|_{*,i}$ the following local ``quasi''-Lipschitz semi-norm:
\begin{equation*}
 % \label{eq:lip1bis}
\|f\|_{*,i} :=  \sup \left\{ \frac{|f(x)-f(y)|}{\theta_{x_i,y_i}+\|x_i-y_i\|} :
 x,y \in\X  \quad \text{$ x_j=y_j$ for $j \neq i$} \right\}  
\end{equation*}
where $\theta_{x,y} = 1$ if $x \neq y$ and $\theta_{x,y} = 0$ otherwise,
for any two elements $x,y \in \X_L$.
This new semi-norm would replace the semi-norm $\|f\|_{1,i}$ defined in
eq.~(\ref{eq:lip1}). Then it is easy to see that all the arguments
carry over also with this new semi-norm and that under this semi-norm
we can approximate uniformly the indicator function of a set $A \in
\A_{\Lambda}$ by the Lipschitz functions
$\Gamma_{A,\rho}(x) = \exp(-\rho^{-1}\inf_{y \in A} |x-y|)$. 
Then, adapting Prop. 2.5 of~\cite{kuensch} we are able to prove that,
under the condition  $\sup_i \sum_{j} C_{ij} |i-j|^{1+\eps} <
\infty$ for some $\eps > 0$ (i.e. $\alpha > 4$) we have
$\alpha_{2,\infty}(n) \le C |n|^{-1-\eps}$ and this is enough to
conclude the proof.  
\end{proof}

% \begin{remark}
% Consider the class of models described by eq.~(\ref{eq:spectral}) with $\omega(k) = |k|$ and $d=3$. Then a sufficient condition for the validity of $(H4)$ with $\alpha > 4$ is
% $$
% \int_{\R^d} \frac{dk}{\omega(k)^{2+\eps}} |k|^2 |\rho(k)|^2 < \infty
% $$  
% for some $\eps > 0$, while for condition $(H3b)$ to hold we must require
% $$
% \int_{\R^d} \frac{dk}{\omega(k)^{1+\eps}} |\rho(k)|^2 < \infty
% $$
% always for some $\eps > 0$, in this case $\gamma = 2+\eps$.
% \end{remark}

\section*{acknowledgments}
  The author would like to thank H.~Spohn and J.~L\H{o}rinczi for
  interesting discussions about the Nelson model and related problems
  which motivated the present work.
The detailed comments of an anonymous referee have contributed to improve the overall quality of the paper.

%%%%%%%%%%%%%%%%%%%%%%%

%\bibliography{gibbs}
%\bibliographystyle{abbrv}

\end{document}